\documentclass[11pt,a4paper,fleqn]{article}
\usepackage{epsfig}
\usepackage{amsmath}
\usepackage{amssymb}
\usepackage{amsfonts}
\usepackage{pifont}
\begin{document}

\title{Cosmic Ray Acceleration by $\bf{ E_{\parallel}}$\\
--- Reconnection of Force-Free Fields}
\author{Stirling A. Colgate\\LANL \and Hui Li\\LANL}
\maketitle {LAUR-04-22-57}
\begin{abstract}
We suggest an unconventional view of the origin of most cosmic rays
(CRs) in the universe.  We propose that nearly every accelerated CR
was part of the parallel current that maintains all force-free (f-f)
magnetic fields.  Charged particles are accelerated by the
$E_{\parallel}$ (to the magnetic filed ${\bf B}$) produced by
reconnection.  The inferred total energy in extra-galactic cosmic rays
is $\sim 10^{60}$ ergs per galaxy spacing volume, provided that
acceleration mechanisms assumed do not preferentially only accelerate
ultra high energy cosmic rays (UHECRs).  This total energy is quite
large, about $\times 10^5$ the parent galactic CR or magnetic energy.
We argue that the formation energy of supermassive black holes
(SMBHs) at galaxy centers, $\sim 10^{62}$ ergs, becomes the only
feasible source. 

We propose an efficient dynamo process which converts gravitational
free energy into magnetic energy in an accretion disk around a
SMBH. Aided by Keplerian winding, this dynamo converts a poloidal seed
field into f-f fields, which are transported into the general
inter-galactic medium (IGM) eventually.  This magnetic energy must
also have been efficiently converted into particle energies, as
evidenced by the radiation from energetic particles. In this view CRs
of the IGM are the result of the continuing dissipation, in a Hubble
time, of this free energy, by acceleration {\it in situ} within the
f-f fields confined within the super-galactic walls and filaments of
large scale structures. In addition, UHECRs are diffusively lost to
the galactic voids {\it before} the GZK attenuation time, $\sim 10^8$
years.  Similarly within the galaxy we expect that the winding by the
disk rotation of the galaxy, by the rotation energy of magnetized
neutron stars, and by the Keplerian winding of star formation disks
are efficient sources of f-f magnetic field energy and hence the
sources of galactic CR acceleration.
\end{abstract}





\section{Introduction}

In this view of the origin of CRs, the source of the necessary free
energy and the form that it takes, force-free magnetic fields, is the
organizational principle in determining the mechanism of CR
acceleration.  Here not only the CR acceleration mechanism is
different from traditional views, but also the implied and necessary
strength and origin of the magnetic fields in the general IGM. There
are literally hundreds of topics and issues to be addressed, both on
the positive and the negative side of this view. We will address a few
of these here, but recognize that attempting to reverse the
conclusions of so many people over so many years is well beyond this
single article. However, we emphasize again that the available free
energy for CR acceleration is the organizing principle of this view of
CR origin.

There are three primary issues:

1) The likely total energy of extra galactic CRs (including radio
lobes) compared to the likely sources of this energy.

2) The astrophysical circumstances for the generation of this energy
as f-f magnetic field energy within the galaxy and within the meta
galaxy.

3) The scaling in time and space of $E_{\parallel}$ reconnection
acceleration leading to the CR spectrum and energy upper limit.

\section{Total Energy in Extra Galactic and Galactic CRs}

\subsection{CR Spectra and Implied Energies}

There are two circumstances where the evidence for extra galactic
cosmic rays appears to be beyond doubt, UHECRs and giant radio
lobes. The first is because the particle energy is great enough,
beyond the "ankle" or $E \ge 10^{18}$ eV, that confinement by the
magnetic fields of the Galaxy is unlikely.  Here the spectral index of
UHECRs returns to nearly the same slope as at lower energies, below
the "knee" at $E \le 10^{15}$ eV. A likely explanation of the change
in slopes, $\Gamma = -2.7, \; -3.0, \; \mbox{and} \; -2.6$, for the
energy regions $ E \le 10^{15}$, $10^{15} \le E \le 10^{18}$, and $\ge
10^{18}$ eV, is that at lower energies, proton escape from the galaxy
at $ E \le 10^{15}$ eV, progressively higher $A/<Z>$ nuclei escape
$\le 10^{15} \le E \le 10^{18}$ eV up to iron at $E \simeq 10^{18}$ eV
and extra galactic protons above this energy. The expected enrichment
at $\le 10^{15} \le E \le 10^{18} $ eV has been observed (Abu-Zayyad
et al. 2001). This spectrum is shown in Figure 1 with various
extrapolated slopes superimposed from $E \le 10^{18}$ down to
$m_pc^2$. Normalizing at $10^{19} $ eV and extrapolating the spectrum
back to $E = m_p c^2$ using a slope of $\Gamma = -2.0$ gives a minimal
estimate of the total energy in extra galactic CRs.  The result is
that the local specific energy content in the extra galactic component
in the meta galaxy, $\epsilon_{mg}$, is $\epsilon_{mg} \simeq 2 \times
10^{-19}$ erg$/cm^3$, as compared to $\epsilon_{gal}
\simeq 10^{-12}$  erg$/cm^3$ for the energy content of the
galactic cosmic ray  spectrum. Since the ratio of the two
volumes is $\sim 4 \;
\mbox{Mp}c^3 / \sim 300 \; \mbox{kpc}^3 \simeq 10^7 $,
both total CR energies, inside and outside the galaxy, would be about
equal. However, such a flat spectrum as $\Gamma = -2.0$ is quite
optimistic for any stochastic acceleration process.  It implies that
the total energy in accelerated particles per logarithmic interval in
energy is a constant so that the integral or total energy diverges
with the upper energy limit.  The back reaction from such an efficient
acceleration mechanism would necessarily alter the source of the
energy leading to a self limiting spectrum. Values of $\Gamma = 0$
describe a modern research accelerator where losses during
acceleration are near zero and the spectrum is a delta function in
energy. The astrophysically reasonable spectrum is for $\Gamma <
-2.0$. Acceleration to such high energies has confounded
astrophysicists for nearly a century.  The slope, $\Gamma = -2.7$, is
constant over 6 decades, for CRs of energy $E_{CR} \le 10^{15}$ eV
within the galaxy and near constant extra galactic over 2 decades with
slope $\Gamma \simeq -2.6$ for $E_{CR} \ge 3 \times 10^{17}$ eV.  The
constant slopes of closely the same value strongly imply a single
scale independent mechanism, so that for extra galactic CRs we expect
the slope $\Gamma \simeq -2.6$ to extend as a constant to low energy
the same as in the galaxy.  The small difference in slopes is likely
explained by the small (over 6 decades) additional nuclear scattering
spallation loss, $\sim 50\%$ in $10^7$ years within the galaxy.

\begin{figure}[htb!]
\begin{center}  
\epsfig{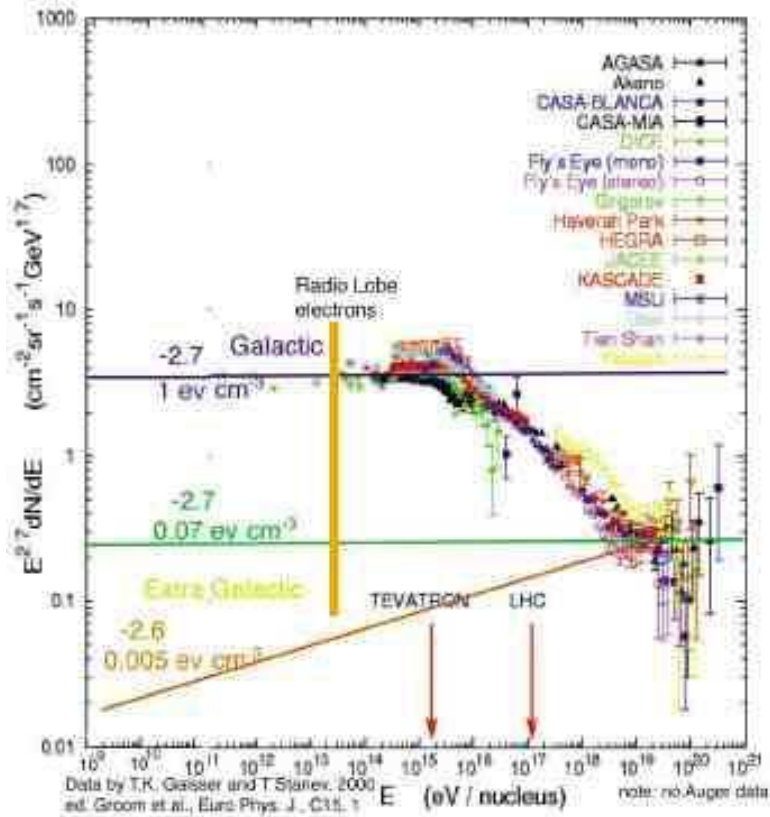}
\end{center}  
\caption{
\small The many measurements of cosmic rays at high energy
as compiled by Gaisser and Stanev (2000). The data are plotted as
$E^{2.7} \, dN/dE$ vs $ E$ and thus with the slope $\Gamma = -2.7$
removed. The change in slope to $\Gamma \simeq -3.0$ in the energy
interval $ E\le 10^{15} \le E \le 10^{18} \ge 10^{18} $ eV is evident
between the "knee" and "ankle". This change in slope is interpreted as
a transition from galactic confinement to extra galactic CRs.  The
width of the transition region is interpreted as progressive loss of
nuclei of progressively higher $<A>/Z$ with the highest being iron
nuclei.  Particles above $10^{18}$ eV are interpreted as
extra galactic UHECRs, primarily protons. Several lines are drawn at
different slopes tangent to the UHECR spectrum and extrapolated back
in energy to $E_{CR} = m_pc^2$ where the maximum in total CR energy
resides. The slope $\Gamma = -2.7$ is the galactic slope and results
in an energy density of extra galactic CRs is $\sim 7 \%$ of the
galactic value. The slope for CRs in the IGM for the same acceleration
mechanism should be slightly flatter because of the lack of
spallation. Spallation attenuates galactic CRs by $\sim 1/2$ in 6
decades in energy and so we estimate that without spallation $\Gamma
\simeq -2.6$.  this results in the energy density of $\simeq 0.005$ of
galactic CRs.  }
\end{figure}

With this assumption in Fig. 1 we have drawn extrapolated spectra
backwards at various slopes from the measured values at high
energy. First we note that at the high energy where the galactic
magnetic field confinement should cease to modify the spectrum, the
"ankle" or $E \ge 10^{18}$ eV, defining CRs in the meta galaxy, the
ratio of energy densities becomes $\epsilon_{mg}/\epsilon_{gal} \simeq
1/30$. This value is comfortably close to what we would expect if
galactic and extra galactic CRs reach equipartition in energy with
what we believe are the magnetic energies within and external to the
galaxy.  We will discuss the origin and evidence for the
extra galactic fields later in more detail.

\subsection{Radio Lobes}

A second indicator of extra galactic cosmic rays is extra galactic
radio lobes.  Giant radio lobes are a strong signature of extra
galactic in-situ particle acceleration.  Since radio lobes are
detected by their radio emission and since strong polarized correlated
emission is observed over distances of up to Mpc, only synchrotron
emission by relativistic electrons in a magnetic field makes a
sensible explanation.  However, it has been known since Burbidge,
(1956) that a minimum in the total energy required to create the radio
lobe emission, (1) wave length, (2) luminosity, and (3) dimension,
occurs for specific values of magnetic field, electron number and
electron energy. These total energies as reported for up to 70 such
giant radio lobes in Kronberg et al. (2001), are immense, up to 10\% of
the MBH rest mass energy. In that analysis the minimum energy is
calculated as if the electrons are accompanied during acceleration by
$\times 100$ the same number of protons at a given energy, just as
observed in the Galaxy. 
However, to be conservative, not knowing the
acceleration mechanism in detail, we assumed, as did others, that the
spectrum commences at the minimum energy necessary of the emitting
electrons, $\gamma = E_e / m_e c^2 \simeq 3\times 10^4$, or $E_p
\simeq 3 \times 10^{13}$ eV with negligible low
energy particles, $E_p/(m_p c^2) \simeq 1$. If we believe that it is
more likely that only one acceleration mechanism prevails for all
energies because of the constancy of $\Gamma$, or equivalently is most
efficient, then extrapolating down to the total energy maximum of the
spectrum, $m_pc^2$ corresponds to an increase of the proton total
energy of $\times 10^{(4.5)^{\Gamma -2}} = 10^{2.47} = \times 300$.
However, the total minimum energy of the lobe increases only as the
$4/7$'s power of the energy of any one component so that minimizing,
but including the low energy extrapolation, increases the minimum
total energy by the factor $\times 10^{(2.47)^{(4/7)}} = 10^{1.4} =
\times 25$ when the new minimum is sought. 
If one chooses not to accelerate the protons, the minimum energy is
reduced by $\sim 10$.  
Thus depending upon assumptions of non-standard
acceleration, the giant radio lobes contain CRs up to a hundred times
the flux of CRs in the IGM or contain only slightly less energy than
the maximum possible back hole formation free energy.  
We conclude that the extra galactic CRs require so much energy, that
only the free energy of black hole formation is a feasible source and
that furthermore the radio lobes are a signature of this energy and
acceleration mechanism. 
 Finally this picture of immense high energy
electron fluxes within radio lobes has recently been confirmed by
Chandra observations of x-rays from high $\gamma \sim 10^3$ electrons
Compton scattering cosmic background photons (Harris 2003).  
We have
concluded separately, Kronberg et al. 2004, that these radio lobes
must be regions of immense f-f fields in quasi magneto hydrostatic
equilibrium.

\subsection{Our Basic Model}

The magnetic field within the Galaxy is well recognized to be $B_{gal}
\simeq 5 \mu G$, but we will claim in this paper the equally
unconventional large magnetic field in the meta galaxy, $B_{mgal}
\simeq 1 \mu G$.  We will discuss further this estimated large value
of the magnetic field in the meta galaxy later, but here note that
this ratio of magnetic energy densities is consistent with the
expected infall pressures, presumably a confinement pressure, of the
IGM accreting onto the galaxy and the infall pressure from the matter
in the voids accreting onto the galaxy walls or filaments. At this
accretion rate the masses of both the galaxy and filaments increase by
$\sim 50\%$ in a Hubble time.  The extra galactic field of $B_{mgal}
\simeq 1 \mu G$ is also consistent with the inferred radio lobe
magnetic fluxes derived from minimum energy derivations, rotation
measure limits, and the theoretical magnetic flux production from the
SMBH accretion disk dynamo. When we slightly increase the extra
galactic index to $\Gamma = -2.6$ from the value $\Gamma = -2.7$
within the galaxy, the total energy density in extra galactic CRs is
then reduced to $\epsilon_{mgal}/\epsilon_{gal} \simeq 1/300$.  This
is sufficient such that the total energy in extra galactic CRs per
galaxy spacing volume of $\sim 4 \; \mbox{Mp}c^3$ becomes $(1/300)
\times 10^{62}$ ergs or $(1/300) \times 10^8 M_{\odot} c^2$, where
$10^8 M_{\odot} c^2$ is the energy available in the accretion disks
that form the SMBHs.  Thus the local, (to the wall or filament) extra
galactic CRs may be replenished $\sim 100$ times in a Hubble time.
This replenishment time, $\sim 10^8$ years, is close to the estimated
loss time by random walk using a semi-coherent intergalactic field of
a galactic spacing (GS) distance, $d_{GS} \simeq 2$ Mpc and a filament
thickness of 5 GS or 5 galaxies thick or 10 Mpc.  The random walk
distance to escape the filament becomes $R_{filament}^2/d_{GS} \simeq
30$ Mpc. This leads to a typical loss time of the CRs to the voids of
$\tau_{loss} \simeq D_{filament}^2/c d_{GS} \simeq 10^8$ years.  This
loss or replenishment time ensures that roughly half of the UHECRs
that are subject to the GZK loss will have been lost to the voids
before detection at Earth provided the total energy available for
continuing acceleration during a Hubble time is $\sim 10^{62}$
ergs. This loss time is also the maximum likely, upper limiting
acceleration time since the f-f fields are primarily confined to the
filaments. The likely upper energy limit of $E_{\parallel}$
acceleration for a particle that remains in a reconnection flux tube
for the entire time will be $E_{max} \simeq B \mu \times \tau_{loss} c
= 3 \times 10^{22}$ eV where $\mu = 300$, the conversion from gauss to
volts at a velocity c. This limit excludes any possible reconnection
current filamentation that may locally increase the acceleration
field, $E_{\parallel}$, but the Larmor radius within the filament and
dependence upon its size and geometry in $10^{-6} \mu$ G field may
limit the maximum energy to an order of magnitude less, $\sim 3 \times
10^{21}$ eV. In this picture there are no local sources, just a
space-filling acceleration and a possible bias in flux opposite to the
direction of the nearest void.

In this picture CRs are accelerated by the same mechanism inside the
Galaxy as well as extra galactic by a space-filling reconnection of
f-f fields; they are closely isotropic, mostly not attenuated by
cosmic background photons, diffuse in an equipartition field both
inside and outside the galaxy, where the flux and field are consistent
with long time mass accretion rate estimates and are consistent with
an energy source of SMBH accretion disks, extra galactic and similar
sources within the Galaxy. We therefore discuss next the basis of f-f
fields in the astrophysical environment and the more limited
possibilities of magnetic field energy generation from the free energy
of SMBH formation.

\section{Force-Free Fields in Astrophysics}

The anzatz of this paper is that once an accretion disk has been
"seeded"  with magnetic flux of sufficient magnitude, then
subsequent accretion will wind up this flux, polloidal, into a
force-free field, a combination of both toroidal and polloidal,
whose total energy is that of the gravitational  energy released
in accretion.  The seed flux is produced by the dynamo
action  occurring either in the MBH accretion disk, or  in
stars (Colgate,  Li \& Pariev 2001).  The physics of the
$\alpha \;\omega$ dynamo is  beyond the subject of this
paper, but without such a dynamo, it is most unlikely that
primordial effects such as the Bierman battery or early phase
transitions  can supply the  seed field  necessary to transfer
the angular momentum of accretion to tension in magnetic field.
The energy contribution of the dynamo is small just as the
exciter field of any commercial electric generator is a
trivial fraction of the power produced, but without an exciter
field no power is produced.  The subsequent flow of power by the
winding of the polloidal field  within a conducting medium,
external to an accretion disk, a Poynting flux, produces the
helical f-f field that we associate with the "jet" of MBHs or jet
identified with star formation (Lovelace , Li et al. 2001).
Fig.2 shows a composite of how  such an accretion disk, dynamo,
and f-f helix are formed. The f-f magnetic helix
is a minimum energy configuration where the major magnetic
stress is maintained by the tension in the field itself.  The
f-f helix is  then self illuminating by the electrons
accelerated by reconnection and  appears as
a jet in shape only.  The relativistic velocities are then a
combination of the
relativistic synchrotron emitting electrons and the phase
velocity of the  reconnection instabilities.

The number density of the current carriers, $n_e$
necessary to carry the current,
$J_{\parallel}$, is trivially small  compared
to the  surrounding IGM, and so the mass  carried by such a jet
is also trivially small.  The "beta" of  relativistic 
current carriers
$v_{drift} \simeq c$ at radii $\sim 3 r_g$, $r_g = MG/c^2$, the
gravitational radius,   is $\beta = n_e \; m_ec^2/(B^2/8\pi)
\sim 10^{-15}$.  Such a f-f field
jet is then  a self illuminated (in x-rays, optical, and radio)
f-f field configuration.  It is illuminated in synchrotron
radiation by   electrons accelerated the same as in the
radio lobes. A similar jet is formed in star formation or by
the winding of flux by either neutron stars, as we believe in
the Crab nebula or by the galaxy itself (not yet observed but
predicted).  In any case the jet  subsequently  morphologically
transforms by  tearing mode reconnection into the radio lobes
and finally into a galactic and  inter galactic space-filling
f-f magnetic flux and therefore a source of free energy.  It is
the continuing reconnection of this space-filling f-f magnetic
flux that we  associate with the production of a near
equipartition flux of galactic and extra galactic  CRs.   Because
of the general lack of  familiarity  with f-f fields in
astrophysics, we give a laboratory example of  the physical
steps necessary to produce such fields and point out the analogy
with accretion disks.  We conclude this section with a
description of the evolution of this flux within cosmic
structure calculations.

\subsection{A Laboratory Example of the Generation of 
Force Free Fields}

The magnetic field inside
a current carrying coil is orthogonal to the boundary current
and the force of the field on the coil winding is $\bf{J
\times B}$ $ = B^2/8\pi$, which we will call a
"force-bounded"  magnetic field.  In particular $\bf {J \cdot
B \simeq  0}$  everywhere  except possibly the coil current
leads.  If now two conducting metal disks  are located  at
either end of the coil and are threaded by the predominantly
axial field, the field will penetrate the conducting metal
disks in a finite time and the field will  return to its
original primarily axial configuration.  If  now the disks
are rotated relative to each other, an electric field will
develop both radial and axial, but nothing further will
happen because the medium between them, air, is an insulator.  If
now the air is replaced by a  conducting fluid medium, e.g.,
plasma or liquid  metal and provided the conductivity, disk
rotation velocity, and dimension are great enough, i.e., the
magnetic Reynolds number, $ Rm = v L/\eta >> 1 $,  a current,
$I_z$ will flow both plus and minus axially at inner and outer
radius. The axial currents are connected to   radial currents,
$I_R$ within the disks. The axial current will produce an
azimuthal field, bounded by the plus and minus axial currents as
well as by the  plus and minus radial currents within the disk.
The vector sum of the axial and azimuthal fields $\bf B_z  +
B_{\phi}$ changes the  topology of the field from initially
axial to helical. Furthermore the radial current in the disks
produce a torque, $R \cdot \bf {J_R \times B_z}$, which, times
the winding of the disks, performs the work that increases the
magnetic energy.   In addition this field will exert a radial
force on the conducting fluid depending upon its confinement, a
"pinch". The same  plasma configuration, a "stabilized pinch" is
produced in the laboratory by replacing the rotation of the
disks  by a voltage (capacitors) between the disks and a low
pressure plasma between them. The resulting current between the
disks  (electrodes) produces the current of the resulting
helical, f-f fields.  If the pressure of the conducting plasma
is small or some of the incompressible liquid metal can escape
either because of instabilities  or through the disk wall, then
the fluid can not exert a force on the field due to the induced
azimuthal field or original axial field.  The resulting
configuration is called  a force-free field  where without
force, ${\bf J \times B  = 0}$ and therefore
$\bf {J = \lambda B}$. In general when the minimum energy
state is reached of such a f-f field, subject to various
boundary conditions of flux and helicity conservation, the
field strength decreases with radius $B \sim \propto 1/R$ so
that a  much weaker outer boundary field exists which must be
supported at some conducting boundary with a weak $\bf{J
\times B} = \nabla (B^2/8\pi) = \nabla P $ force. This boundary
pressure in the astrophysical case is the pressure of the IGM
or ISM. However,  the main point is that the differential
rotation of the disks  has added free energy to the field
by the work done by twisting the field, (equivalently supplied by
the capacitor in laboratory plasma experiments) thereby
increasing both components of the field,
$(B_z^2 + B_{\phi}^2)/8\pi$. Furthermore the
major fraction of the field  pressure is supported by the tension in
the field itself  and so is called force-free. This energy can  be
accessed by
allowing the disks to unwind due to the torque of the tension
in the field, an electric motor, or more likely because of
reconnection dissipated by $\bf{E \cdot J}$   or
$\eta \bf{J^2}$.

\subsection{Formation of Astrophysical Force-Free Fields
and Their Distribution }

In astrophysics the winding of accretion disks forming nearly
every compact object plays the role of the conducting disks.
The original angular momentum of matter, both
baryonic and dark is formed randomly by the three body
interactions in the first non-linear gravitational collapse
starting from initial small perturbations (Peebles 1969).  The
specific  angular momentum is large, $\sim \times 10^7$ greater
than the limiting  specific angular momentum of say  the MBH of
every galaxy. Subsequently a mass selection, $M_{MBH}\simeq
10^{-3}M_{gal}$,  and angular
momentum transport process, the Rossby vortex mechanism (Li et
al. 2001) are invoked to explain how such an accretion disk
could form (Colgate et al. 2003). A similar angular momentum is
formed in pre-stellar structure formation from molecular clouds,
cores, in star formation, but here the ratio of the specific
angular momentum of the core to the limiting specific angular
momentum of  the star is less,
$\sim  300$.
Regardless, once an accretion disk is initiated, half the
gravitational energy of forming the collapsed object must be
released  in the transport of angular momentum of
the matter accreted to the condensed object.  By far the largest
free energy  is released in forming the SMBH, $10^8 M_{\odot} c^2
= 10^{62}$  ergs versus
$10^{11} M_{\odot} \epsilon_{stars} = 10^{59}$ ergs where
$\epsilon_{stars} \simeq 10^{15}$ ergs/g, the specific binding energy of
the average star.  Hence we concentrate on the the MBH case where
we see a direct mechanism for converting this free energy
into f-f magnetic field energy.

Fig.2  shows the sequence of all three processes  leading
to the generation of the inter galactic  f-f fields. The
pre-collapse from the Lyman-$\alpha$ cloud to the flat rotation
curve galaxy is not shown, but a mass selection is made at a
critical thickness by the Rossby vortex mechanism  leading to a
disk of $\sim 10^8 M_{\odot}$  (Colgate et
al.  2003) in which
angular momentum is transported by the co-rotating Rossby
vortices shown as  the disk with vortices (Li et al., 2001).
Star-disk collisions produce the helicity to produce an  $\alpha
- \omega$  dynamo supplying the polloidal, seed
field.  The differential winding of the Keplerian disk flow
produces the f-f helix.  The f-f helix extends away from the
disk a distance  determined by winding number, $n_w \simeq 10^8 \,
\mbox{years}/(2 \pi r_g /c)\simeq 10^{11}$ turns,
or $\sim 10$ Mpc. Reconnection shortens this distance  to radio
lobe sizes of a Mpc.

\begin{figure}[htb!]
\begin{center}  
\epsfig{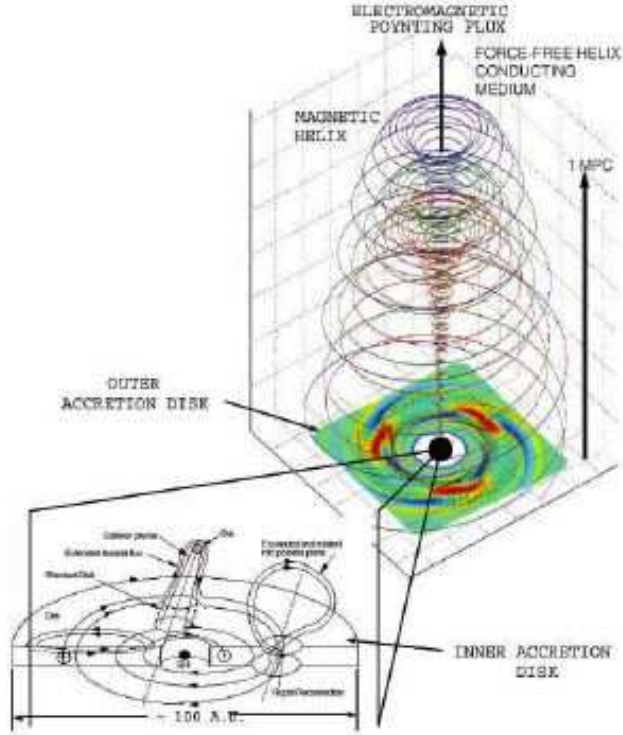}
\end{center}  
\caption{
\small A concept of the formation of  massive back
holes in galaxies. A galaxy forms as a "flat rotation curve"
disk, $M \propto R$.  At some radius  in this mass distribution
the column  density becomes great enough, $\Sigma_{crit} \simeq
100$ g$cm^{-3}$ such that the Rossby vortex instability is
excited, shown as the co-rotating vortices in the disk.  The
resulting transport of angular momentum  allows the interior mass
to collapse forming the black hole.  This simple criterion
predicts the mass and mass-velocity correlation of MBHs. An
$\alpha \omega$ dynamo forms within  this disk  because of the
differential rotation and the helicity injected by star
collisions with the disk.  The dynamo, with near infinite gain
in  $n_t \simeq10^{12}$ revolutions in $\sim 10^8$ years,
provides the (small) polloidal flux out of which the helix is
formed by the differential winding of the foot prints.  This
helix is force-free  because  the field lines are  primarily
axial and azimuthal and so  matter, tied to the field lines,
falls back to the disk in the strong gravity.  The  helix
extends a distance of $D \simeq n_t 2 \pi R_g \simeq 60$ Mpc.
Reconnection leads to radio lobes at smaller distances.
}
\end{figure}

\subsection{Filling the filaments with magnetic flux}

The life time of the radio lobes is roughly the same as the
formation time of the MBH,  $\sim 10^8$  years, and so in a
Hubble time, 100 times longer, we expect the fields to evolve
and fill the IGM quasi uniformly.  A simulation of the structure
of a local region of the universe with magnetic fields  has been
made by Ryu, Kang, \& Bierman (1998) in a $32\; \times\;$ 32 Mpc
box, shown in Fig. 3. Galaxies  with typical galaxy spacings are
superimposed.  In this calculation a primordial magnetic field
was  assumed  and its  strength  chosen such as to allow the
evolution of structure and the support of this structure as
typically observed in  filaments, (walls) and voids and with a
modulation in density of $\sim 20:1$.   The variations in the
orientation of the fields are due to the instabilities  inherent
to gravitational structure formation. However, what is more
important is the bounding pressure of the filaments which is
derived in the calculation from the infall  of matter from the
voids.  This infall is the fundamental instability of structure
formation due to gravity and dark matter.  The dark matter
follows its own  gravitational perturbations and continually
passes back and forth through itself.  The baryons first follow
this gravitational perturbation but then  can not pass through
themselves, and therefore  make shocks  and pressure.  This
pressure is usually assumed to support the filaments from
collapse, but cooling should reduce this   pressure leading to
further collapse of the filaments.  Instead, in these
calculations  the magnetic pressure supports the filaments both
against further collapse, but more importantly against the
pressure of gas continuing to infall from the voids.  It is
this  infall pressure that confines the magnetic field. They
found that the equilibrium magnetic field pressure that  retains
the necessary density modulation ratio of  filament density  to
void  density had  to be as  large as $\sim 1 \mu$ G.  This is
the field  shown  in Fig 3. This is very closely the magnetic
field  strength expected if
$10\%$ of the MBH formation energy,
$10^{61}$ ergs is distributed in a GS box, $Vol \simeq 10^{74}$
$cm^3$.  This value of the IGM field is not  inconsistent with the
background Faraday rotation measures of distant polarized
sources, provided one averages field reversals on a scale much
smaller, $\sim 1/10 \; \mbox{to} \;  1/100 $ of GS, uses an
electron density of $n_e \sim  10^{-5}$
$cm^{-3}$ and includes the volume ratio of  filaments to voids.
We expect the synchrotron glow from these distributed, weak
fields  will outline the structures of the universe when such
low  frequency arrays as LOFAR  are operational.   For now, only
the radio lobes and their minimum energy indicate
consistency with this picture of the magnetic free energy of the
universe.  With this view of the dominant free energy in the
universe in the form of f-f  fields, we proceed to an
abbreviated theory of reconnection acceleration of CRs.

\begin{figure}[htb!]
\begin{center}  
\epsfig{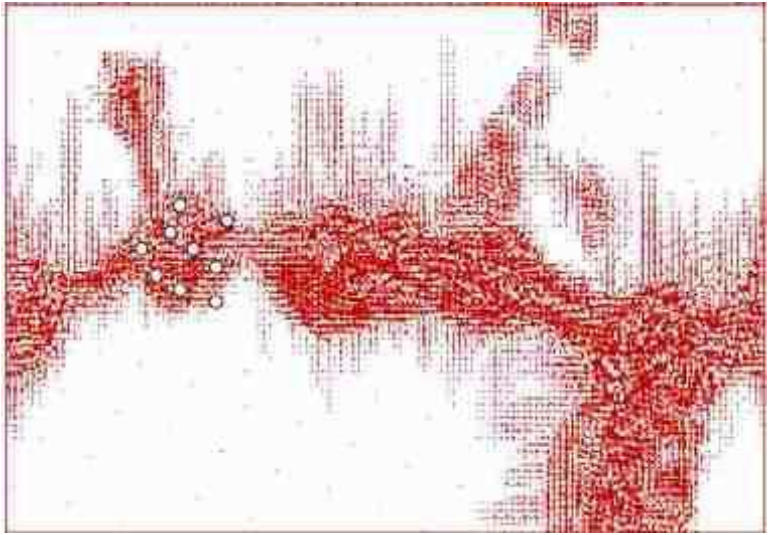}
\end{center}  
\caption{
\small A cosmological  structure simulation $32 \times 32$
Mp$c^3$, by Ryu, Kang, and Bierman 1998, in which magnetic field
was tied to the matter and the structure  evolved with magnetic
pressure as well  as gravity.  The final, near-equilibrium
state,  was one where the magnetic field of  $\sim 1 \mu $ G was
confined to the  filaments by the  in-fall pressure of matter
from the  voids.  In our model of a magnetized universe the
magnetic flux, instead of being assumed and adjusted, is instead
produced in excess by the winding of the accretion disks and
reaches the steady state defined by the in-fall pressure and
reconnection. Hence we have superimposed on a local region of a
filament, an approximate distribution   of symbolic galaxies.
CRs will diffuse between galaxies following lines of force until
escaping to the voids in
$\sim 10^8$ years. This avoids observation of most of the CRs
that might have been attenuated by the  GZK mechanism.
}
\end{figure}

\section{Reconnection Acceleration of CRs}


\subsection{Dissipation of Force-Free Fields in the Laboratory
and Astrophysics}

By way of comparison to shock acceleration, there is a vast literature
on the experimental observation and its interpretation of dissipation,
reconnection of, and particle acceleration in f-f fields. The
organizing physical principle is the maximization of the dissipation
rate of the magnetic free energy. The free energy of a f-f magnetic
field can be accessed or converted to another form, heat or kinetic
energy, only by ${\bf E \cdot J}$. A resistive origin of $E = \eta J$ 
leads to heat, but the accleration of relativistic  run-away current
carriers of say energy $\gamma_{i,e} m_{i,e} c^2$, leads to a kinetic
energy density  $n_{J,i,e} m_{i,e} \gamma_{i,e} c^2$ where  the number
density of current carriers is $n_{J,i,e} = J/ec$.  In order for
the current of run-away carriers  to exceed the usual or
expected much larger  slow drift of typical current carrying background
plasma, and thereby effectively cause $E_{\parallel}  \sim 0$, some
un-identified instability  must locally  immobilize these background
plasma charged particles so that due to current  carrier starvation a
large $E_{\parallel}$  is maintained, 
$E_{accel} \propto dJ/dt$.  It is reasonable that the greater rigidity
of the run-away or accelerated current carriers  should circumvent
the impedance of such an immobilization instability, but it is
puzzling why, as we observe in cosmic rays, that  the bulk of the
particles  energy centers around $c^2$  or $\gamma \simeq 1$.

In tokamaks, reconnection sometimes leads to "current
interuptions" where all the free energy of the magnetic field is
transformed into a run-away beam of relativistic ions or electrons
that sometimes "melts a hole" in the metal vacuum liner.  Most often
reconnection leads to an interchange of flux surfaces and a
consequential loss of confinement of the hot plasma to the cold walls,
a multi billion dollar question for fusion. In stabilized pinches or
reverse field pinches, more analogous to the above described helical
fields and to astrophysical fields where the winding number is large,
and so where $B_{\phi} \simeq B_{z}$, the flux surface topology becomes
tangled and random.  An experimental demonstration of these tangled
fields is given in the appendix of Colgate (1978).  We suspect, but
cannot prove, that the maximum dissipation rate occurs when the
current carriers are accelerated to a velocity, $\sim c$, or $\gamma
\sim 1$. The velocity from further acceleration is then bounded, 
and some optimum  rigidity occurs for a given acceleration. High
rigidity particles should remain in a reconnecting flux filament the
longest, thereby gaining the largest energy from the
$E_{\parallel}$ of reconnection. The momentum of the current carriers,
$\gamma m_i c$, then inertially carries the current following the tangled
field lines on average to a larger radius and thereby rapidly diffuses
the current in radius.  This is a very complicated non-linear sequence
of energy flow and plasma and field deformations, but it contains the
necessary phenomena of acceleration, diffusion and dissipation. It has
not yet been simulated.

\subsection{Particle acceleration by magnetic reconnection}

As mentioned before, radio lobes are clear examples where CRs are
being accelerated in a magnetized IGM. To understand particle
acceleration by reconnection in lobes, it is instructive to begin in
the resistive MHD limit, even though it is probably {\em not} valid
for radio lobes, given their large size and small resistivity $--$ in
which case ordinary magnetic field diffusion will not be fast enough
to account for the magnetic energy conversion. For example, in a
filament of size of $\sim 1$ kpc, and resistivity of $\eta \sim 10^4$
cm$^2$/s (using an electron temperature of $10^6$ K), the diffusion
time will be $L^2/\eta \sim 9 \times 10^{38}$ sec, much longer than
the Hubble time. A very different situation obtains, however, by
realizing that as the fluids carry the frozen-in fields and move them
around, steep field gradients could be generated; These result in thin
sheet-like current structures, and hence greatly reduce the diffusion
times. In the Sweet-Parker reconnection picture, in which the current
layer width is $\Delta_{\eta} \sim (\tau_A \eta)^{1/2}$, where $\tau_A
\sim L/v_A$, the typical MHD time-scale, and $v_A \sim 6.6\times 10^8
B_{3.10^{-6}}/n_{-6}^{1/2}$ cm/s. Then the rate of energy dissipation
is related to the rate of convection of magnetic flux into and out of
the reconnection region. This time scale (again in the Sweet-Parker
model) is $\tau_{sp} \sim (\tau_A \tau_\eta)^{1/2} \sim 6\times
10^{25}$ sec, but clearly still much too long to be relevant to radio
lobes.

The physical conditions of the lobes are, rather, more consistent
with the so-called fast collisionless reconnection scenario, which
has recently been studied in the context of hot fusion laboratory
plasmas (e.g., tokamaks) and magnetospheric plasmas (e.g., Earth's
magnetotail). This is because for radio lobes, the ion skin depth
(which is understood to be closely related to kinetic effects in
reconnection), $d_i = c/\omega_{pi} \sim 2.3\times 10^{10}
n_{-6}^{-1/2}$ cm, is actually larger than the resistive
Sweet-Parker layer width
\begin{equation}
\label{eq:sw_layer}
\Delta_{sp} \sim 2.1\times 10^8
\left(\frac{L}{1 {\rm kpc}}\right)^{1/2}
\left(\frac{3\times 10^{-6} {\rm G}}{B}\right)^{1/2}
\left(\frac{n}{10^{-6}}\right)^{1/4}
\left(\frac{\eta}{10^4}\right)^{1/2} ~~{\rm cm}~~.
\end{equation} In this limit, reconnection is mediated by the
kinetic physics to break the flux frozen-in condition. It has been
suggested and shown that the reconnection rate is then independent
of the resistivity (e.g., Shay \& Drake 1998). The exact dependence
of the reconnection rate on various parameters (especially $d_i$),
however, is under debate (Shay et al. 1999; Wang et al. 2001;
Fitzpatrick 2003). Under a simplified geometry and in 2D, the latter
two studies suggested that the length of the current layer
undergoing reconnection depends on the boundary driving, which is
unfortunately very difficult to determine in the radio lobe
situation. Another recent study by Li et al. (2003) on a fully
force-free system using particle-in-cell simulations has shown that
collisionless reconnection that is facilitated by the full kinetic
physics can indeed proceed at a very fast rate, with flow speeds
close to a fraction of the Alfv\'en speed.

The above scale estimates strongly favor the idea that collisionless
reconnection in radio lobes will be Alfv\'enic, and could play an
important role in converting the magnetic energy to particles at a
fast rate, given the high Alfv\'{e}n speeds within radio lobes. This
may therefore be the main mechanism of in situ particle acceleration,
as demanded by the radio spectral index distributions. More detailed
scaling studies, especially the dependence on the system size, need to
be done before we can model the role of reconnection in radio lobes in
more detail.

For the general IGM, a similar comparison between the resistive layer
width and the ion skin depth can be made. The higher density and
likely lower magnetic field strength will tend to bring these two
scales closer but the key uncertainty will be whether and how thin
current sheets can be produced via, say, ideal MHD processes. Many
uncertainties remain, such as the global magnetic field configurations
in the lobes and the general IGM, how efficiently thin current sheets
can be made, etc.

\section{The Power-law Spectrum}

The accepted theory of cosmic ray acceleration is shock
wave acceleration in the ISM driven by supernova (Axford,
Leer, \& Skadron, 1977; Bell, 1978; Blandford \&
Ostriker, 1988).  ``This acceptance has been largely based
upon the good agreement between the "universal" power-law
spectrum predicted by shock acceleration, i.e., the
power-law index becomes:

	$$ \Gamma = (d \;lnN)/(d\; ln E) \simeq -(2+\epsilon) $$

depending only on the Mach number and the observed or
inferred particle spectra." (see Blandford \& Eichler,
1987 for a review, and many papers by P. Biermann for a
more accurate comparison.)  This belief that a nearly
correct power-law spectral index alone is unique is
instead, a less restrictive condition than commonly
believed. Any accelerator for which a fractional gain in
energy, $d \; ln E$, by a few particles is accompanied by
a  fractional loss, $-d \; lnN$, in number of the
remainder will give a power-law:

	$$ dN/N = - \Gamma (dE/E) ~~.$$

The fractional loss for a fractional gain in energy is
what would be expected for a rigidity dependent loss
mechanism where the probability of a relativistic particle
being scattered out of an acceleration region is inversely
proportional to its energy or rigidity.
	For values of $\Gamma \ge \sim -2$, i.e. a smaller
fractional loss, the integral energy becomes
asymptotically large,   and at some energy will truncate
or limit the acceleration mechanism, destroying the
confinement and hence the accelerating mechanism itself.
Hence, it is not likely that at any one time we should see
many such accelerators occurring naturally in the Galaxy.
On the other hand accelerators with $\Gamma << - 2$ will
produce a steep spectrum that, relative to another less
steep one, i.e. $\Gamma$ closer to -2, will be lost
relative to the less steep mechanisms above some critical
energy. Hence it is likely that the spectrum of any
observed mechanism should be close to $ \Gamma= -(2+\epsilon)$.

\section{Experiments}

Laboratory experiments can be performed to simulate both
magneto-hydrodynamics as well as the tearing mode reconnection and the
associated $E_{\parallel}$ acceleration of the "run-away" particles.
The spheromak and reverse-field pinch experiments are a step in this
direction.  Interruptions in tokamaks are already laboratory proof of
this acceleration. Experimental proof of an $\alpha-\Omega$ dynamo is
similarly needed. We need to perform more laboratory plasma
experiments to observe reconnection in the collisionless limit.  Without
laboratory experiments, as for example shock acceleration, we are
still uncertain about the origin of cosmic rays.

Acknowledgements: We are indebted to Philipp Kronberg for extensive
knowledge and analysis of radio lobes and to many other colleagues in
support of the project of a magnetized universe. This research was
performed under the auspices of the Department of Energy. It was
supported by the Laboratory Directed Research and Development Program
at Los Alamos.  S.A.C and H.L. acknowledge the hospitality of Aspen
Center for Physics while part of the research was carried out during
the summers of 2001 \& 2002.


\begin{thebibliography}{000}




\bibitem{}Abu-Zayyad, T.,  et al., 2001,  ApJ. 557, 686

\bibitem{}Axford, W.I., Leer, E., \& Skadron, G., 1977, 15th
ICRC, 11, 132

\bibitem{}Bell, A.R., 1978, MNRAS, 182,147

\bibitem{}Blandford, R.D. \& Eichler, D., 1987, Physics Reports,
154, 1, 1987

\bibitem{}Blandford, R.D. \& Ostriker, J.P., 1988, ApJ, 221, L29

\bibitem{}Burbidge, G.R. 1956, ApJ, 124, 416

\bibitem{}Colgate, S.A. \& Li, H., 1999,  Astrophys. Space Sci. ,
264. 357

\bibitem{}Colgate, S.A. \& Li, H., 2000,  IAU Symp 195, ASP Conf.
Series 334, eds.  P.C.H. Martens and S. Tsurta

\bibitem{}Colgate, S.A., Li, H., Pariev, V.I., 2001, Physics of
Plasmas, 8, 2425

\bibitem{}Colgate, S.A., 1978, Ap. J., 221, 1068 .

\bibitem{}Colgate, S.A., Li, H., Pariev, V.I., 2001, Physics of
Plasmas, 8, 2425

\bibitem{colgateetal} S.~A.~Colgate R.~ Cen, N.~Currier, H.~Li, M.~S.~Warren, 
ApJ., {\bf 558}, ( 2003) L7. 

\bibitem{}Gaisser, T.K.  and Stanev, T., 2000,  European Physical Journal C
vol. 15, pp. 150-156.

\bibitem{}Harris, D., 2003, New Astronomy, 47, 617

\bibitem{}Fitzpatrick, R. 2003, Phys Plasmas 10, 1702

\bibitem{}Jokipii, J.R., \& Morfill, G., 1987, ApJ. 312, 170

\bibitem{} Choudhuri, A. \&  Konigl, A., 1986,   ApJ 310, 96.

\bibitem{}Kronberg, P.P., Dufton, Q.W., Li, H., \& Colgate, S.A. 2001,
ApJ  560, 178

\bibitem{}Li, H., Colgate, S.A., Wendroff, B., \&
Liska, R. 2001, ApJ 551, 874

\bibitem{}Li, H., Lovelace, R.V.E., Finn, J.M., and Colgate, S.A. 2001,
ApJ, 561, 915

\bibitem{}Li, H. et al. 2003, Phys Plasmas, 10, 2763

\bibitem{}Lovelace, R.V.E. 1976, {\it Nature}\, 262, 649

\bibitem{}Peebles, P.J.E. 1969, ApJ, 155, 393

\bibitem{}Ryu, D., Kang, H., \& Bierman, P.L., 1998, Astron. Astrophys.
335, 19.

\bibitem{} Shay, M.A., \& Drake, J.F. 1998, Geophys. Rev. Lett.,
25, 3759

\bibitem{}Shay, M.A., Drake, J.F., \& Rogers, B.N. 1999,
Geophys. Rev. Lett., 26, 2163

\bibitem{}Wang, X., Bhattacharjee, A., \& Ma, Z.W. 2001,
PRL, 87, 265003

\end{thebibliography}
\end{document}